\documentclass[a4paper]{PoS}

\usepackage{amsmath}
\usepackage[utf8]{inputenc}
\usepackage{nicefrac}
\usepackage{subfigure}
\usepackage{tikz}

\title{The Rho Resonance from $N_f=2+1+1$ Twisted Mass Lattice QCD}

\ShortTitle{The $\rho$ resonance from $N_f=2+1+1$ tmLQCD}

\author{C.~Helmes, C.~Jost, B.~Knippschild, L.~Liu, C.~Urbach, \speaker{M.~Werner}\\
  HISKP (Theory), University of Bonn, Nussallee 14-16, Bonn, Germany\\
  E-mail: \email{werner@hiskp.uni-bonn.de}}

\author{Z.~Wang\\
  School of Physics, Peking University, Beijing, China}

\author{for the European Twisted Mass Collaboration}

\abstract{We present first results on the $\rho$ resonance parameters
  obtained with $N_f=2+1+1$ Wilson twisted mass fermions at maximal
  twist. Using ensembles of the ETM collaboration, we provide results
  for two values of the lattice spacing and a range of pion mass
  values.}

\FullConference{The 33rd International Symposium on Lattice Field Theory\\
		14 -18 July 2015\\
		Kobe International Conference Center, Kobe, Japan*}

\begin{document}

\section{Introduction}

\begin{table}[t]
  \begin{center}
    \begin{tabular}{lccccc}
      \hline\hline
      name  \quad & $L/a$ \quad \quad & $T/a$ \quad \quad & $N_\mathrm{conf}$ \quad & $a[\text{fm}]$ \quad & $M_{\pi}[\text{MeV}]$\\
      \hline\hline
      A30.32 & 32 & 64 & 147 & 0.086 & 147 \\ 
      A40.32 & 32 & 64 & 250 & 0.086 & 250 \\ 
      A60.24 & 24 & 48 & 245 & 0.086 & 246\\ 
      A80.24 & 24 & 48 & 251 & 0.086 & 254\\ 
      B55.32 & 32 & 64 & 146 & 0.082 & 146\\
      \hline\hline
    \end{tabular}
  \end{center}
  \caption{All Ensembles are generated by the European Twisted Mass 
    Collaboration with 
    $N_f=2+1+1$ Wilson twisted mass fermions \cite{Baron:2010bv,Baron:2011sf}}
  \label{tab:etmcenssembles}
\end{table}

The $\rho$ resonance is located in the isospin-1 channel of $\pi\pi$
scattering. The $\rho$ decays strongly almost
exclusively into $\pi\pi$. It is besides the $f_0$ or $\sigma$
resonance the lowest resonance in QCD. Therefore, it is highly
desirable to gain theoretical insight using a non-perturbative method
like lattice QCD. 

Resonance properties are not directly accessible in lattice QCD
due to the Euclidean structure of space-time~\cite{Maiani:1990ca}. Martin
L{\"u}scher invented an approach utilising finite size effects in
energy levels of multi-particle states, which can be directly related to
the infinite volume scattering phase shift.

In this proceeding we present preliminary results of the first
calculation of the $\rho$ resonance properties using lattice QCD with
$N_f=2+1+1$ dynamical quark flavours. The computation is based on
gauge configurations generated by the European Twisted Mass
Collaboration (ETMC)~\cite{Baron:2010bv,Baron:2011sf} and uses the
Wilson twisted 
mass discretisation of QCD~\cite{Frezzotti:2000nk}, which has
the property of automatic $\mathcal{O}(a)$
improvement~\cite{Frezzotti:2003ni} at
so-called maximal twist. Details of the gauge
configurations used including the spatial and temporal lattice extends
$L/a$ and $T/a$, respectively, and the number of configurations
$N_\mathrm{conf}$ can be found in table~\ref{tab:etmcenssembles}. For
more details we refer to the corresponding references from which we also
took over the notation. For other results on the $\rho$ meson from
this conference see Refs.~\cite{Bulava:2015qjz,Guo:2015dde}.

\section{L{\"u}scher Method}

The relation between energy levels in a finite volume and the 
infinite volume phase shift was first described by L{\"u}scher in
Refs.~\cite{Luscher:1985dn,Luscher:1986pf,Luscher:1990ux,Luscher:1991cf}.
It was later on extended and generalised for moving
frames~\cite{Rummukainen:1995vs,Feng:2010es}. In this paper we follow
the notations of Ref.~\cite{Gockeler:2012yj}. In a general moving
frame with momentum $\vec P  = 2\pi\vec n/L$, one first computes the
energy $E$ of the $\pi\pi$ system with $I=1$ for the interacting case
using suitable operators as described in the next section. 
Next, the center-of-mass frame energy 
\[
E_\mathrm{CM}^2 = E^2 - \vec P^2
\]
and the corresponding Lorentz factor
\[
\gamma = \frac{E}{E_\mathrm{CM}}
\] 
are computed.
From the center-of-mass energy the lattice-momentum transfer fraction,  
$q$, can be derived via
\[
\tilde q^2 = \frac{E_\mathrm{CM}^2}{4}-M_\pi^2\,,\qquad q^2 =
\left(\frac{\tilde q L}{2\pi}\right)^2\,,
\]
where $M_\pi$ is the pion rest mass. With the definition of $\gamma$
and $q$ it is possible to connect the finite-volume energies with phase shift values
via
\begin{equation}
  \label{eq:ws}
  w_{lm} =
  \frac{1}{\pi^{3/2}\sqrt{2l+1}} \gamma^{-1}q^{-l-1}
  \mathcal{Z}_{lm}^{\vec n}(q^2)\,. 
\end{equation}
where $\mathcal{Z}_{lm}^{\vec n}(q^2)$ is the generalised L{\"u}scher
$\mathcal{Z}$-function.
For example, the phase shift in the
center of mass frame ($\vec n =0, \gamma=1$), $\delta_1$, is obtained  via~\cite{Luscher:1991cf}
\begin{equation}
  \label{eq:delta}
  \cot\delta_1\ =\ w_{00}\,.
\end{equation}
Similar relations hold
also for moving frames, see e.g. Ref.~\cite{Gockeler:2012yj}. 

In principle mixing with higher partial waves needs to be taken into
account but it has been shown in Ref.~\cite{Dudek:2012xn} that the
mixing is negligible in practice. We will assume this holds for
our data at the moment. Nevertheless, we will investigate 
the validity of this assumption in the future.

The experimental data for the scattering phase shift
$\delta_1(E_\mathrm{CM})$ can be described using the effective range
formula~\cite{Brown:1968zza}
\begin{equation}
  \label{eq:ERF}
  \tan\delta_1(E_\mathrm{CM})\ =\ \frac{g_{\rho\pi\pi}^2}{6\pi}
  \frac{\tilde q^3}{E_\mathrm{CM}(M_\rho^2-E_\mathrm{CM}^2)}\,,
\end{equation}
with two parameters, the effective coupling $g_{\rho\pi\pi}$ and the
$\rho$ meson mass $M_\rho$, respectively. Following
Ref.~\cite{Feng:2010es}, we are going to use this formula to
describe our lattice data for the scattering phase shift
$\delta_1(E_\mathrm{CM})$. $M_\rho$ and $g_{\rho\pi\pi}$ are related
to the $\rho$ meson decay width $\Gamma_\rho$ by 
\begin{equation}
  \label{eq:coupling}
  \Gamma_\rho\ =\ \frac{g_{\rho\pi\pi}^2 \left(\nicefrac{M_\rho^2}{4}
      - M_\pi^2\right)^{\frac{3}{2}}}{6 \pi M_\rho^2}\,,
\end{equation}
with $M_\pi$ the pion mass.

\section{Operators and Data Analysis}

\begin{figure}[t]
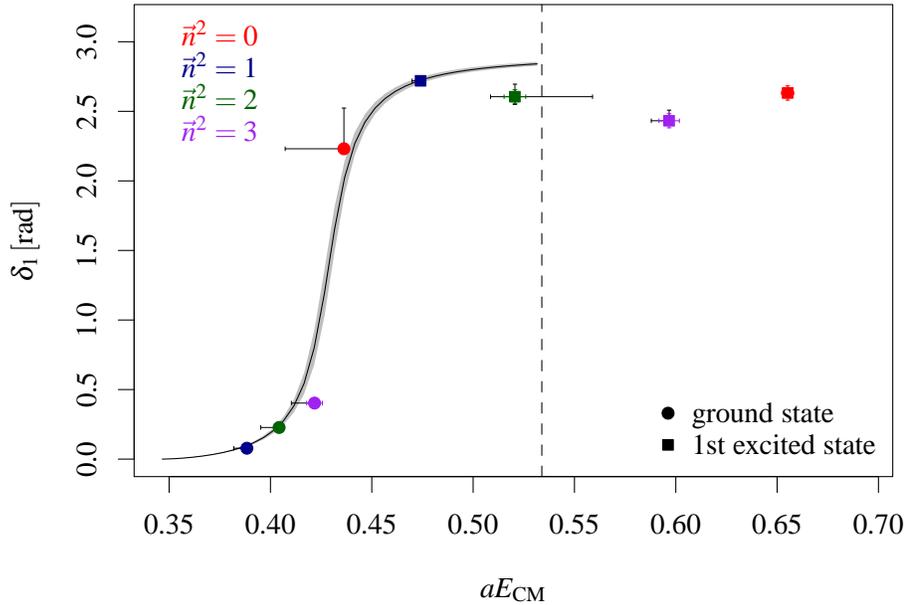

  \begin{center}
    \include{delta-A60.24}
    \caption{$I=1$ $\pi\pi$ scattering phase shift for the A60
      ensemble. The solid line is a fit of the effective range
      formula Eq.~\protect\ref{eq:ERF} 
      to our data with statistical errors indicated by the band. The vertical line
      indicates the $2M_K$ threshold.}
    \label{fig:A40resonance}
  \end{center}
\end{figure}

We determine the interacting energy levels $E$ by computing the
following $2\times2$ correlation matrix 
\begin{equation}
  \label{eq:C}
  \mathcal{C}(\vec P, t) = 
  \langle\,\vec O\ \otimes\ \vec O^\dagger\,\rangle\,,\qquad \vec O =
  (\pi\pi(\vec P, t), \rho^0(\vec P, t))^t\,.
\end{equation}
The interpolating operators $\pi\pi(\vec P, t)$ and $\rho^0(\vec P, t)$
are the two particle and single particle operators, respectively,
which couple to the isospin-1 channel depending on the total center of
mass momentum $\vec P$. For details on the construction of these operators see
e.g. Refs.~\cite{Feng:2010es,Dudek:2012xn}. For this proceeding we
concentrate on the $A_1$ irreducible representation of the octahedral
group~\cite{Gockeler:2012yj}. We include all frames with $\vec n^2 =
0,1,2,3$, including all possible permutations contribution to the same $\vec n^2$.

As a smearing scheme we use the stochastic Laplacian Heaviside method
as described in Refs.~\cite{Peardon:2009gh,Morningstar:2011ka}. The
details and our parameter choices can be found in
Ref.~\cite{Helmes:2015gla}.

Note that in Wilson twisted mass lattice QCD isospin is broken
explicitly at $\mathcal{O}(a^2)$ which leads to fermionic disconnected
contributions to the $\rho^0$. These
contributions are pure lattice artefacts and will, therefore, be
neglected in our analysis, see also Ref.~\cite{Feng:2010es}. 

Next we solve the generalised eigenvalue problem
\[
\mathcal{C}(t)\ \eta^{(n)}(t,t_0) = \lambda^{(n)}(t,t_0)\
\mathcal{C}(t_0)\ \eta^{(n)}(t,t_0)
\]
for eigenvectors $\eta^{(n)}(t,t_0)$ and eigenvalues
$\lambda^{(n)}(t,t_0)$,
$n=0,1$~\cite{Michael:1982gb,Luscher:1990ck}. The energy levels are
determined from the exponential fall-off of $\lambda(t,t_0)$ at large
$t$-values. Each $2\times2$ correlation matrix allows us to access two
energy levels.
We take the so-called thermal pollutions~\cite{Feng:2009ij} into
account by weighting and shifting of
$\mathcal{C}$~\cite{Dudek:2012xn}. 

The analysis follows the one we already described in
Ref.~\cite{Helmes:2015gla}.
We use a bootstrap procedure with $5000$ bootstrap samples to compute
statistical errors and to estimate the variance-covariance matrix used
in the $\chi^2$-fits. For a given eigenvalue we fit an exponential
function with two free parameters to the data for a large set of fit
ranges $[t_1,t_2]$ with degrees of freedom larger than $5$ for the
lowest energy level and larger than $4$ for the first excited
state. The p-value of each fit indicates whether or not this model is
justified. The energy level is, therefore, determined as the weighted
median over all used fit ranges with weight~\cite{Helmes:2015gla} 
\[
w_\mathrm{E} = \{(1-2|p_\mathrm{E} - 0.5|)
\cdot1/\Delta_\mathrm{E}\}^2\,, 
\]
where $p_\mathrm{E}$ is the p-value of the corresponding exponential
fit and $\Delta_\mathrm{E}$ the statistical error of the energy level
determined from the bootstrap procedure. Similarly, we determine the
pion mass $M_\pi$ and derived quantities like $\delta_1$, see
Ref.~\cite{Helmes:2015gla}. In the computation of $\delta_1$ we
exclude energy levels where $E_\mathrm{CM} < 2 M_\pi$, which might
appear due to 
statistical fluctuations. A systematic uncertainty on $E$ and
$\delta_1$ stemming from the 
fit ranges is obtained from the $68\%$ confidence interval of the
corresponding weighted distributions. 

\section{Results}

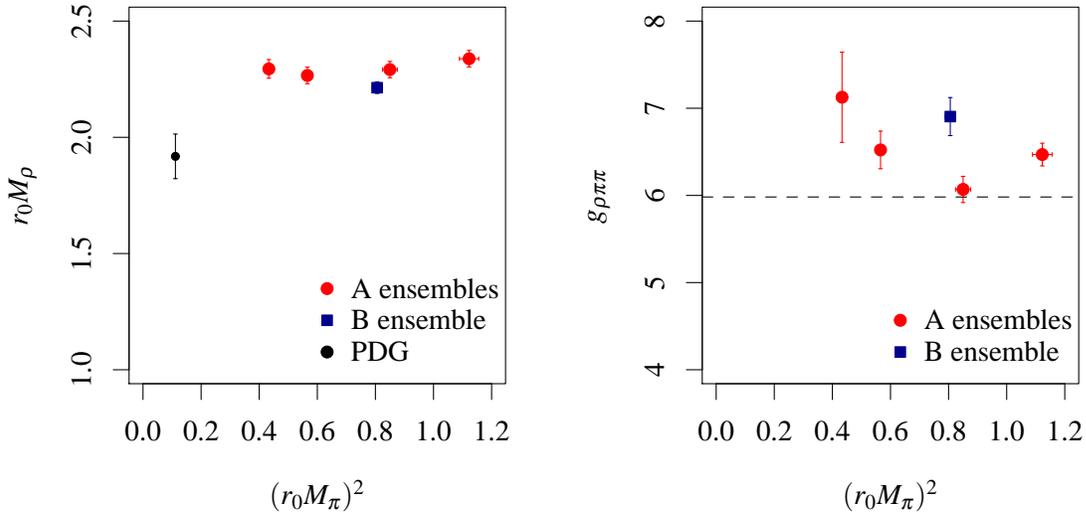
\begin{figure}[t]
  \begin{center}
\begin{tikzpicture}[x=1pt,y=1pt]
\definecolor[named]{fillColor}{rgb}{1.00,1.00,1.00}
\path[use as bounding box,fill=fillColor,fill opacity=0.00] (0,0) rectangle (216.81,252.94);
\begin{scope}
\path[clip] (  0.00,  0.00) rectangle (216.81,252.94);
\definecolor[named]{drawColor}{rgb}{0.00,0.00,0.00}

\path[draw=drawColor,line width= 0.4pt,line join=round,line cap=round] ( 54.47, 61.20) -- (186.34, 61.20);

\path[draw=drawColor,line width= 0.4pt,line join=round,line cap=round] ( 54.47, 61.20) -- ( 54.47, 55.20);

\path[draw=drawColor,line width= 0.4pt,line join=round,line cap=round] ( 76.45, 61.20) -- ( 76.45, 55.20);

\path[draw=drawColor,line width= 0.4pt,line join=round,line cap=round] ( 98.43, 61.20) -- ( 98.43, 55.20);

\path[draw=drawColor,line width= 0.4pt,line join=round,line cap=round] (120.41, 61.20) -- (120.41, 55.20);

\path[draw=drawColor,line width= 0.4pt,line join=round,line cap=round] (142.38, 61.20) -- (142.38, 55.20);

\path[draw=drawColor,line width= 0.4pt,line join=round,line cap=round] (164.36, 61.20) -- (164.36, 55.20);

\path[draw=drawColor,line width= 0.4pt,line join=round,line cap=round] (186.34, 61.20) -- (186.34, 55.20);

\node[text=drawColor,anchor=base,inner sep=0pt, outer sep=0pt, scale=  1.00] at ( 54.47, 39.60) {0.0};

\node[text=drawColor,anchor=base,inner sep=0pt, outer sep=0pt, scale=  1.00] at ( 76.45, 39.60) {0.2};

\node[text=drawColor,anchor=base,inner sep=0pt, outer sep=0pt, scale=  1.00] at ( 98.43, 39.60) {0.4};

\node[text=drawColor,anchor=base,inner sep=0pt, outer sep=0pt, scale=  1.00] at (120.41, 39.60) {0.6};

\node[text=drawColor,anchor=base,inner sep=0pt, outer sep=0pt, scale=  1.00] at (142.38, 39.60) {0.8};

\node[text=drawColor,anchor=base,inner sep=0pt, outer sep=0pt, scale=  1.00] at (164.36, 39.60) {1.0};

\node[text=drawColor,anchor=base,inner sep=0pt, outer sep=0pt, scale=  1.00] at (186.34, 39.60) {1.2};

\path[draw=drawColor,line width= 0.4pt,line join=round,line cap=round] ( 49.20, 66.48) -- ( 49.20,198.47);

\path[draw=drawColor,line width= 0.4pt,line join=round,line cap=round] ( 49.20, 66.48) -- ( 43.20, 66.48);

\path[draw=drawColor,line width= 0.4pt,line join=round,line cap=round] ( 49.20,110.47) -- ( 43.20,110.47);

\path[draw=drawColor,line width= 0.4pt,line join=round,line cap=round] ( 49.20,154.47) -- ( 43.20,154.47);

\path[draw=drawColor,line width= 0.4pt,line join=round,line cap=round] ( 49.20,198.47) -- ( 43.20,198.47);

\node[text=drawColor,rotate= 90.00,anchor=base,inner sep=0pt, outer sep=0pt, scale=  1.00] at ( 34.80, 66.48) {1.0};

\node[text=drawColor,rotate= 90.00,anchor=base,inner sep=0pt, outer sep=0pt, scale=  1.00] at ( 34.80,110.47) {1.5};

\node[text=drawColor,rotate= 90.00,anchor=base,inner sep=0pt, outer sep=0pt, scale=  1.00] at ( 34.80,154.47) {2.0};

\node[text=drawColor,rotate= 90.00,anchor=base,inner sep=0pt, outer sep=0pt, scale=  1.00] at ( 34.80,198.47) {2.5};

\path[draw=drawColor,line width= 0.4pt,line join=round,line cap=round] ( 49.20, 61.20) --
	(191.61, 61.20) --
	(191.61,203.75) --
	( 49.20,203.75) --
	( 49.20, 61.20);
\end{scope}
\begin{scope}
\path[clip] (  0.00,  0.00) rectangle (216.81,252.94);
\definecolor[named]{drawColor}{rgb}{0.00,0.00,0.00}

\node[text=drawColor,anchor=base,inner sep=0pt, outer sep=0pt, scale=  1.00] at (120.41, 15.60) {$(r_0M_\pi)^2$};

\node[text=drawColor,rotate= 90.00,anchor=base,inner sep=0pt, outer sep=0pt, scale=  1.00] at ( 10.80,132.47) {$r_0 M_\rho$};
\end{scope}
\begin{scope}
\path[clip] ( 49.20, 61.20) rectangle (191.61,203.75);
\definecolor[named]{drawColor}{rgb}{1.00,0.00,0.00}
\definecolor[named]{fillColor}{rgb}{1.00,0.00,0.00}

\path[draw=drawColor,line width= 0.4pt,line join=round,line cap=round,fill=fillColor] (116.65,177.88) circle (  2.25);

\path[draw=drawColor,line width= 0.4pt,line join=round,line cap=round] (116.65,174.73) -- (116.65,181.03);

\path[draw=drawColor,line width= 0.4pt,line join=round,line cap=round] (115.93,174.73) --
	(116.65,174.73) --
	(117.37,174.73);

\path[draw=drawColor,line width= 0.4pt,line join=round,line cap=round] (117.37,181.03) --
	(116.65,181.03) --
	(115.93,181.03);

\path[draw=drawColor,line width= 0.4pt,line join=round,line cap=round] (114.75,177.88) -- (118.55,177.88);

\path[draw=drawColor,line width= 0.4pt,line join=round,line cap=round] (114.75,178.61) --
	(114.75,177.88) --
	(114.75,177.16);

\path[draw=drawColor,line width= 0.4pt,line join=round,line cap=round] (118.55,177.16) --
	(118.55,177.88) --
	(118.55,178.61);

\path[draw=drawColor,line width= 0.4pt,line join=round,line cap=round,fill=fillColor] (102.08,180.37) circle (  2.25);

\path[draw=drawColor,line width= 0.4pt,line join=round,line cap=round] (102.08,176.83) -- (102.08,183.91);

\path[draw=drawColor,line width= 0.4pt,line join=round,line cap=round] (101.36,176.83) --
	(102.08,176.83) --
	(102.80,176.83);

\path[draw=drawColor,line width= 0.4pt,line join=round,line cap=round] (102.80,183.91) --
	(102.08,183.91) --
	(101.36,183.91);

\path[draw=drawColor,line width= 0.4pt,line join=round,line cap=round] (100.64,180.37) -- (103.52,180.37);

\path[draw=drawColor,line width= 0.4pt,line join=round,line cap=round] (100.64,181.10) --
	(100.64,180.37) --
	(100.64,179.65);

\path[draw=drawColor,line width= 0.4pt,line join=round,line cap=round] (103.52,179.65) --
	(103.52,180.37) --
	(103.52,181.10);

\path[draw=drawColor,line width= 0.4pt,line join=round,line cap=round,fill=fillColor] (147.87,180.11) circle (  2.25);

\path[draw=drawColor,line width= 0.4pt,line join=round,line cap=round] (147.87,177.01) -- (147.87,183.20);

\path[draw=drawColor,line width= 0.4pt,line join=round,line cap=round] (147.15,177.01) --
	(147.87,177.01) --
	(148.59,177.01);

\path[draw=drawColor,line width= 0.4pt,line join=round,line cap=round] (148.59,183.20) --
	(147.87,183.20) --
	(147.15,183.20);

\path[draw=drawColor,line width= 0.4pt,line join=round,line cap=round] (145.08,180.11) -- (150.67,180.11);

\path[draw=drawColor,line width= 0.4pt,line join=round,line cap=round] (145.08,180.83) --
	(145.08,180.11) --
	(145.08,179.38);

\path[draw=drawColor,line width= 0.4pt,line join=round,line cap=round] (150.67,179.38) --
	(150.67,180.11) --
	(150.67,180.83);

\path[draw=drawColor,line width= 0.4pt,line join=round,line cap=round,fill=fillColor] (177.82,184.21) circle (  2.25);

\path[draw=drawColor,line width= 0.4pt,line join=round,line cap=round] (177.82,181.07) -- (177.82,187.36);

\path[draw=drawColor,line width= 0.4pt,line join=round,line cap=round] (177.10,181.07) --
	(177.82,181.07) --
	(178.54,181.07);

\path[draw=drawColor,line width= 0.4pt,line join=round,line cap=round] (178.54,187.36) --
	(177.82,187.36) --
	(177.10,187.36);

\path[draw=drawColor,line width= 0.4pt,line join=round,line cap=round] (174.18,184.21) -- (181.46,184.21);

\path[draw=drawColor,line width= 0.4pt,line join=round,line cap=round] (174.18,184.93) --
	(174.18,184.21) --
	(174.18,183.49);

\path[draw=drawColor,line width= 0.4pt,line join=round,line cap=round] (181.46,183.49) --
	(181.46,184.21) --
	(181.46,184.93);
\definecolor[named]{drawColor}{rgb}{0.00,0.00,0.55}
\definecolor[named]{fillColor}{rgb}{0.00,0.00,0.55}

\path[draw=drawColor,line width= 0.4pt,line join=round,line cap=round,fill=fillColor] (141.03,171.26) rectangle (145.02,175.24);

\path[draw=drawColor,line width= 0.4pt,line join=round,line cap=round] (143.02,171.09) -- (143.02,175.41);

\path[draw=drawColor,line width= 0.4pt,line join=round,line cap=round] (142.30,171.09) --
	(143.02,171.09) --
	(143.75,171.09);

\path[draw=drawColor,line width= 0.4pt,line join=round,line cap=round] (143.75,175.41) --
	(143.02,175.41) --
	(142.30,175.41);

\path[draw=drawColor,line width= 0.4pt,line join=round,line cap=round] (141.20,173.25) -- (144.85,173.25);

\path[draw=drawColor,line width= 0.4pt,line join=round,line cap=round] (141.20,173.97) --
	(141.20,173.25) --
	(141.20,172.53);

\path[draw=drawColor,line width= 0.4pt,line join=round,line cap=round] (144.85,172.53) --
	(144.85,173.25) --
	(144.85,173.97);
\definecolor[named]{drawColor}{rgb}{0.00,0.00,0.00}
\definecolor[named]{fillColor}{rgb}{0.00,0.00,0.00}

\path[draw=drawColor,line width= 0.4pt,line join=round,line cap=round,fill=fillColor] ( 66.74,147.25) circle (  1.50);

\path[draw=drawColor,line width= 0.4pt,line join=round,line cap=round] ( 66.74,138.81) -- ( 66.74,155.69);

\path[draw=drawColor,line width= 0.4pt,line join=round,line cap=round] ( 66.02,138.81) --
	( 66.74,138.81) --
	( 67.46,138.81);

\path[draw=drawColor,line width= 0.4pt,line join=round,line cap=round] ( 67.46,155.69) --
	( 66.74,155.69) --
	( 66.02,155.69);
\definecolor[named]{drawColor}{rgb}{1.00,0.00,0.00}
\definecolor[named]{fillColor}{rgb}{1.00,0.00,0.00}

\path[draw=drawColor,line width= 0.4pt,line join=round,line cap=round,fill=fillColor] (124.12, 97.20) circle (  2.25);
\definecolor[named]{drawColor}{rgb}{0.00,0.00,0.55}
\definecolor[named]{fillColor}{rgb}{0.00,0.00,0.55}

\path[draw=drawColor,line width= 0.4pt,line join=round,line cap=round,fill=fillColor] (122.13, 83.21) rectangle (126.12, 87.19);
\definecolor[named]{drawColor}{rgb}{0.00,0.00,0.00}
\definecolor[named]{fillColor}{rgb}{0.00,0.00,0.00}

\path[draw=drawColor,line width= 0.4pt,line join=round,line cap=round,fill=fillColor] (124.12, 73.20) circle (  2.25);

\node[text=drawColor,anchor=base west,inner sep=0pt, outer sep=0pt, scale=  1.00] at (133.12, 93.76) {A ensembles};

\node[text=drawColor,anchor=base west,inner sep=0pt, outer sep=0pt, scale=  1.00] at (133.12, 81.76) {B ensemble};

\node[text=drawColor,anchor=base west,inner sep=0pt, outer sep=0pt, scale=  1.00] at (133.12, 69.76) {PDG};
\end{scope}
\end{tikzpicture}\unskip
\begin{tikzpicture}[x=1pt,y=1pt]
\definecolor[named]{fillColor}{rgb}{1.00,1.00,1.00}
\path[use as bounding box,fill=fillColor,fill opacity=0.00] (0,0) rectangle (216.81,252.94);
\begin{scope}
\path[clip] (  0.00,  0.00) rectangle (216.81,252.94);
\definecolor[named]{drawColor}{rgb}{0.00,0.00,0.00}

\path[draw=drawColor,line width= 0.4pt,line join=round,line cap=round] ( 54.47, 61.20) -- (186.34, 61.20);

\path[draw=drawColor,line width= 0.4pt,line join=round,line cap=round] ( 54.47, 61.20) -- ( 54.47, 55.20);

\path[draw=drawColor,line width= 0.4pt,line join=round,line cap=round] ( 76.45, 61.20) -- ( 76.45, 55.20);

\path[draw=drawColor,line width= 0.4pt,line join=round,line cap=round] ( 98.43, 61.20) -- ( 98.43, 55.20);

\path[draw=drawColor,line width= 0.4pt,line join=round,line cap=round] (120.41, 61.20) -- (120.41, 55.20);

\path[draw=drawColor,line width= 0.4pt,line join=round,line cap=round] (142.38, 61.20) -- (142.38, 55.20);

\path[draw=drawColor,line width= 0.4pt,line join=round,line cap=round] (164.36, 61.20) -- (164.36, 55.20);

\path[draw=drawColor,line width= 0.4pt,line join=round,line cap=round] (186.34, 61.20) -- (186.34, 55.20);

\node[text=drawColor,anchor=base,inner sep=0pt, outer sep=0pt, scale=  1.00] at ( 54.47, 39.60) {0.0};

\node[text=drawColor,anchor=base,inner sep=0pt, outer sep=0pt, scale=  1.00] at ( 76.45, 39.60) {0.2};

\node[text=drawColor,anchor=base,inner sep=0pt, outer sep=0pt, scale=  1.00] at ( 98.43, 39.60) {0.4};

\node[text=drawColor,anchor=base,inner sep=0pt, outer sep=0pt, scale=  1.00] at (120.41, 39.60) {0.6};

\node[text=drawColor,anchor=base,inner sep=0pt, outer sep=0pt, scale=  1.00] at (142.38, 39.60) {0.8};

\node[text=drawColor,anchor=base,inner sep=0pt, outer sep=0pt, scale=  1.00] at (164.36, 39.60) {1.0};

\node[text=drawColor,anchor=base,inner sep=0pt, outer sep=0pt, scale=  1.00] at (186.34, 39.60) {1.2};

\path[draw=drawColor,line width= 0.4pt,line join=round,line cap=round] ( 49.20, 66.48) -- ( 49.20,198.47);

\path[draw=drawColor,line width= 0.4pt,line join=round,line cap=round] ( 49.20, 66.48) -- ( 43.20, 66.48);

\path[draw=drawColor,line width= 0.4pt,line join=round,line cap=round] ( 49.20, 99.48) -- ( 43.20, 99.48);

\path[draw=drawColor,line width= 0.4pt,line join=round,line cap=round] ( 49.20,132.47) -- ( 43.20,132.47);

\path[draw=drawColor,line width= 0.4pt,line join=round,line cap=round] ( 49.20,165.47) -- ( 43.20,165.47);

\path[draw=drawColor,line width= 0.4pt,line join=round,line cap=round] ( 49.20,198.47) -- ( 43.20,198.47);

\node[text=drawColor,rotate= 90.00,anchor=base,inner sep=0pt, outer sep=0pt, scale=  1.00] at ( 34.80, 66.48) {4};

\node[text=drawColor,rotate= 90.00,anchor=base,inner sep=0pt, outer sep=0pt, scale=  1.00] at ( 34.80, 99.48) {5};

\node[text=drawColor,rotate= 90.00,anchor=base,inner sep=0pt, outer sep=0pt, scale=  1.00] at ( 34.80,132.47) {6};

\node[text=drawColor,rotate= 90.00,anchor=base,inner sep=0pt, outer sep=0pt, scale=  1.00] at ( 34.80,165.47) {7};

\node[text=drawColor,rotate= 90.00,anchor=base,inner sep=0pt, outer sep=0pt, scale=  1.00] at ( 34.80,198.47) {8};

\path[draw=drawColor,line width= 0.4pt,line join=round,line cap=round] ( 49.20, 61.20) --
	(191.61, 61.20) --
	(191.61,203.75) --
	( 49.20,203.75) --
	( 49.20, 61.20);
\end{scope}
\begin{scope}
\path[clip] (  0.00,  0.00) rectangle (216.81,252.94);
\definecolor[named]{drawColor}{rgb}{0.00,0.00,0.00}

\node[text=drawColor,anchor=base,inner sep=0pt, outer sep=0pt, scale=  1.00] at (120.41, 15.60) {$(r_0M_\pi)^2$};

\node[text=drawColor,rotate= 90.00,anchor=base,inner sep=0pt, outer sep=0pt, scale=  1.00] at ( 10.80,132.47) {$g_{\rho\pi\pi}$};
\end{scope}
\begin{scope}
\path[clip] ( 49.20, 61.20) rectangle (191.61,203.75);
\definecolor[named]{drawColor}{rgb}{1.00,0.00,0.00}
\definecolor[named]{fillColor}{rgb}{1.00,0.00,0.00}

\path[draw=drawColor,line width= 0.4pt,line join=round,line cap=round,fill=fillColor] (116.65,149.69) circle (  2.25);

\path[draw=drawColor,line width= 0.4pt,line join=round,line cap=round] (116.65,142.57) -- (116.65,156.81);

\path[draw=drawColor,line width= 0.4pt,line join=round,line cap=round] (115.93,142.57) --
	(116.65,142.57) --
	(117.37,142.57);

\path[draw=drawColor,line width= 0.4pt,line join=round,line cap=round] (117.37,156.81) --
	(116.65,156.81) --
	(115.93,156.81);

\path[draw=drawColor,line width= 0.4pt,line join=round,line cap=round] (114.74,149.69) -- (118.56,149.69);

\path[draw=drawColor,line width= 0.4pt,line join=round,line cap=round] (114.74,150.41) --
	(114.74,149.69) --
	(114.74,148.97);

\path[draw=drawColor,line width= 0.4pt,line join=round,line cap=round] (118.56,148.97) --
	(118.56,149.69) --
	(118.56,150.41);

\path[draw=drawColor,line width= 0.4pt,line join=round,line cap=round,fill=fillColor] (102.08,169.64) circle (  2.25);

\path[draw=drawColor,line width= 0.4pt,line join=round,line cap=round] (102.08,152.56) -- (102.08,186.71);

\path[draw=drawColor,line width= 0.4pt,line join=round,line cap=round] (101.36,152.56) --
	(102.08,152.56) --
	(102.80,152.56);

\path[draw=drawColor,line width= 0.4pt,line join=round,line cap=round] (102.80,186.71) --
	(102.08,186.71) --
	(101.36,186.71);

\path[draw=drawColor,line width= 0.4pt,line join=round,line cap=round] (100.62,169.64) -- (103.54,169.64);

\path[draw=drawColor,line width= 0.4pt,line join=round,line cap=round] (100.62,170.36) --
	(100.62,169.64) --
	(100.62,168.91);

\path[draw=drawColor,line width= 0.4pt,line join=round,line cap=round] (103.54,168.91) --
	(103.54,169.64) --
	(103.54,170.36);

\path[draw=drawColor,line width= 0.4pt,line join=round,line cap=round,fill=fillColor] (147.87,134.75) circle (  2.25);

\path[draw=drawColor,line width= 0.4pt,line join=round,line cap=round] (147.87,129.79) -- (147.87,139.70);

\path[draw=drawColor,line width= 0.4pt,line join=round,line cap=round] (147.15,129.79) --
	(147.87,129.79) --
	(148.59,129.79);

\path[draw=drawColor,line width= 0.4pt,line join=round,line cap=round] (148.59,139.70) --
	(147.87,139.70) --
	(147.15,139.70);

\path[draw=drawColor,line width= 0.4pt,line join=round,line cap=round] (145.01,134.75) -- (150.74,134.75);

\path[draw=drawColor,line width= 0.4pt,line join=round,line cap=round] (145.01,135.47) --
	(145.01,134.75) --
	(145.01,134.02);

\path[draw=drawColor,line width= 0.4pt,line join=round,line cap=round] (150.74,134.02) --
	(150.74,134.75) --
	(150.74,135.47);

\path[draw=drawColor,line width= 0.4pt,line join=round,line cap=round,fill=fillColor] (177.82,147.94) circle (  2.25);

\path[draw=drawColor,line width= 0.4pt,line join=round,line cap=round] (177.82,143.65) -- (177.82,152.24);

\path[draw=drawColor,line width= 0.4pt,line join=round,line cap=round] (177.10,143.65) --
	(177.82,143.65) --
	(178.54,143.65);

\path[draw=drawColor,line width= 0.4pt,line join=round,line cap=round] (178.54,152.24) --
	(177.82,152.24) --
	(177.10,152.24);

\path[draw=drawColor,line width= 0.4pt,line join=round,line cap=round] (174.11,147.94) -- (181.53,147.94);

\path[draw=drawColor,line width= 0.4pt,line join=round,line cap=round] (174.11,148.67) --
	(174.11,147.94) --
	(174.11,147.22);

\path[draw=drawColor,line width= 0.4pt,line join=round,line cap=round] (181.53,147.22) --
	(181.53,147.94) --
	(181.53,148.67);
\definecolor[named]{drawColor}{rgb}{0.00,0.00,0.55}
\definecolor[named]{fillColor}{rgb}{0.00,0.00,0.55}

\path[draw=drawColor,line width= 0.4pt,line join=round,line cap=round,fill=fillColor] (141.03,160.33) rectangle (145.02,164.31);

\path[draw=drawColor,line width= 0.4pt,line join=round,line cap=round] (143.02,155.11) -- (143.02,169.52);

\path[draw=drawColor,line width= 0.4pt,line join=round,line cap=round] (142.30,155.11) --
	(143.02,155.11) --
	(143.75,155.11);

\path[draw=drawColor,line width= 0.4pt,line join=round,line cap=round] (143.75,169.52) --
	(143.02,169.52) --
	(142.30,169.52);

\path[draw=drawColor,line width= 0.4pt,line join=round,line cap=round] (141.14,162.32) -- (144.90,162.32);

\path[draw=drawColor,line width= 0.4pt,line join=round,line cap=round] (141.14,163.04) --
	(141.14,162.32) --
	(141.14,161.60);

\path[draw=drawColor,line width= 0.4pt,line join=round,line cap=round] (144.90,161.60) --
	(144.90,162.32) --
	(144.90,163.04);
\definecolor[named]{drawColor}{rgb}{0.00,0.00,0.00}

\path[draw=drawColor,line width= 0.4pt,dash pattern=on 4pt off 4pt ,line join=round,line cap=round] ( 49.20,131.88) -- (191.61,131.88);
\definecolor[named]{drawColor}{rgb}{1.00,0.00,0.00}
\definecolor[named]{fillColor}{rgb}{1.00,0.00,0.00}

\path[draw=drawColor,line width= 0.4pt,line join=round,line cap=round,fill=fillColor] (124.12, 85.20) circle (  2.25);
\definecolor[named]{drawColor}{rgb}{0.00,0.00,0.55}
\definecolor[named]{fillColor}{rgb}{0.00,0.00,0.55}

\path[draw=drawColor,line width= 0.4pt,line join=round,line cap=round,fill=fillColor] (122.13, 71.21) rectangle (126.12, 75.19);
\definecolor[named]{drawColor}{rgb}{0.00,0.00,0.00}

\node[text=drawColor,anchor=base west,inner sep=0pt, outer sep=0pt, scale=  1.00] at (133.12, 81.76) {A ensembles};

\node[text=drawColor,anchor=base west,inner sep=0pt, outer sep=0pt, scale=  1.00] at (133.12, 69.76) {B ensemble};
\end{scope}
\end{tikzpicture}
    \caption{Left: $\rho$ resonance mass $r_0 M_\rho$ as a function of
      the squared pion mass $(r_0 M_\pi)^2$ for all ensembles investigated
      in this work. The black point is the
      corresponding PDG value~\cite{Agashe:2014kda}. Right: $g_{\rho\pi\pi}$
      versus the squared 
      pion mass $(r_0 M_\pi)^2$. The dashed horizontal line corresponds to the
      PDG value.}
    \label{fig:chiralmrho}
  \end{center}
\end{figure}

In figure~\ref{fig:A40resonance} we show the P-wave scattering phase
shift as a function of $aE_\mathrm{CM}$ for the A60.24 ensemble
(cf. Table~\ref{tab:etmcenssembles}) below inelastic threshold.
We show statistical errors as the inner error bars and the systematic
error as the outer error bars in the figure. As visible, the
systematic uncertainties are sizable in some cases which is a hint
that we do not yet control excited or thermal states satisfactorily
with the weighting and shifting procedure. In the future we will
increase the operator basis to get a better handle on these effects.

We observe that -- as expected for a resonance -- the phase-shift
starts around $0$, rises sharply through $\pi/2$ around
$aE_\mathrm{CM}=0.43$ and levels out again somewhat below
$3$. The fit of Eq.~\ref{eq:ERF} to the data is shown
in figure~\ref{fig:A40resonance} as the solid line with error
band. Statistical errors on $E_\mathrm{CM}$ and $\delta_1$ are
included correlated in the fit, while systematic uncertainties are not
yet included.
All points in the range from $2M_\pi$ to $2M_K$ are included
in the fit.
The fit describes the data well, though the $\chi^2$ value is not
particularly good. This comes mainly from the decrease in $\delta_1$
at $E_\mathrm{CM}$ close to $2M_K$. We observe a similar behaviour on
all our ensembles. The large $\chi^2$-value might also be due to not
completely removed excited or thermal states, which we are
currently investigating. Note, however, that the $\chi^2$
value reduces already significantly once systematic uncertainties are
included in the fit.

We have carried out this analysis for all ensembles listed in
table~\ref{tab:etmcenssembles}. We remark that the ensembles B55.32
and A30.32 currently have significantly lower statistics than the
other three ensembles.
The resulting values of $M_\rho$ are shown in units
of the Sommer parameter $r_0$ in  the left plot of figure~\ref{fig:chiralmrho} as a
function of $(r_0 M_\pi)^2$. The values and errors of $r_0/a$ are
taken from Ref.~\cite{Carrasco:2014cwa}. The errors of
$r_0$ are propagated via parametric bootstrap to $r_0M_\rho$.
Additionally, we show the PDG value of $M_\rho=
775.5\,\mathrm{MeV}$~\cite{Agashe:2014kda} using $r_0=0.49(2)\
\mathrm{fm}$. The error on this point is solely coming from the
uncertainty in the physical value of $r_0$.
In the right plot of figure~\ref{fig:chiralmrho} we show $g_{\rho\pi\pi}$ again as a
function of $(r_0 M_\pi)^2$. The uncertainties of $g_{\rho\pi\pi}$ are
significantly larger than for $M_\rho$ due to the structure of the ERF
Eq.~\ref{eq:ERF}. 
The experimental value of $g_{\rho\pi\pi}$ included in the figure was
obtained by inserting $M_\rho=775.5\,\mathrm{MeV}$, $\Gamma_\rho=
149.1 \,\mathrm{MeV}$ and $M_\pi= 139.57 \,\mathrm{MeV}$ into
Eq.~\ref{eq:coupling}. Within the currently still sizable
uncertainties for the effective coupling we can confirm the mild
dependence on $M_\pi^2$ observed in previous lattice investigations. 
Note that we did not yet estimate a systematic
uncertainty for $M_\rho$ and $g_{\rho\pi\pi}$.

\section{Summary and Outlook}

We have presented first results of the $\rho$ meson resonance
properties obtained with $N_f=2+1+1$ Wilson twisted mass fermions. The
computations have been performed for two values of the lattice
spacing and a range of pion mass values. 

For a number of ETMC ensembles we have determined the scattering phase
shift $\delta_1$ as a function of the center of mass energy. Using the
effective range formula, we obtain the $\rho$ resonance mass $M_\rho$
and the effective coupling $g_{\rho\pi\pi}$. This study will be
extended by enlarging the operator basis and by 
including more irreducible representations which will help to 
investigate systematic errors stemming from excited or thermal states
further. To analyse lattice artifacts and perform a chiral extrapolation 
we will increase the number of lattice spacing values and pion masses.

We thank the
members of ETMC for the most enjoyable collaboration. The computer
time for this project was made available to us by the John von
Neumann-Institute for Computing (NIC) on the JUDGE and Juqueen
systems in J{\"u}lich. 
We thank U.-G.~Mei{\ss}ner for granting us 
access on JUDGE. This project was funded by the DFG as a project in
the Sino-German CRC110. Z. Wang was supported in part by the
National Science Foundation of China (NSFC) under the project
 o.11335001. The open source software
packages tmLQCD~\cite{Jansen:2009xp}, Lemon~\cite{Deuzeman:2011wz} and
R~\cite{R:2005} have been used.

\bibliographystyle{h-physrev5}
\bibliography{bibliography}

\end{document}